\documentclass[prl,twocolumn,tightlines,showpacs,preprintnumbers,aps,floatfix,showpacs,amsmath,amssymb]{revtex4}
\usepackage{epsfig,amsmath,amssymb,bm,epsf,graphics}
\usepackage{dcolumn}
\newcommand{\beq}{\begin{equation}}
\newcommand{\beqa}{\begin{eqnarray}}
\newcommand{\eeq}{\end{equation}}
\newcommand{\eeqa}{\end{eqnarray}}
\renewcommand{\a}{\alpha}

\renewcommand{\d}{{\rm d}}
\newcommand{\de}{_{(2)}}

\newcommand\dt[1]{\frad{\d#1}{\d t}}
\newcommand{\e}{{\rm e}}

\newcommand{\frad}[2]{\displaystyle{\displaystyle#1\over\displaystyle#2}}
\newcommand{\g}{\gamma}
\newcommand{\gbar}{\overline{g}}
\newcommand{\infy}{_{(\infty)}}
\newcommand{\infymax}{_{(\infty){\rm max}}}

\renewcommand{\max}{_{\rm max}}
\newcommand{\m}{{\bf m}}

\newcommand{\meansur}[1]{\langle\!\langle#1\rangle\!\rangle}
\newcommand{\n}{{\bf n}}
\newcommand{\bl}{$\bullet$}
\newcommand{\s}{\sigma}
\newcommand{\one}{_{(1)}}
\newcommand{\vi}{$\circ$}

\newcommand{\N}{{\cal N}}

\begin{document}

\title{From ageing to  immortality: cluster growth in stirred colloidal solutions}

\author{Anita Mehta }
\email{anita@bose.res.in}
\affiliation{S. N. Bose National Centre for Basic Sciences, Block JD, Salt Lake, Kolkata 700 098, India}

\date{\today}

\begin{abstract}
Our model describes cluster aggregation in a stirred colloidal solution
Interacting clusters compete for growth in this 'winner-takes-all'
model; for finite assemblies, the largest cluster always wins, i.e.
there is a uniform sediment.
In mean-field, the model exhibits
glassy dynamics, with two well-separated time scales,
corresponding to individual and collective behaviour; the survival probability of a cluster
eventually falls off according to a universal law $(\ln t)^{-1/2}$.
In finite dimensions, the glassiness is enhanced:
the dynamics  manifests 
 both {\it ageing} and metastability, where pattern formation is manifested
in each metastable state by a fraction of {\it immortal} clusters.
\end{abstract}

\pacs{68.43.Jk, 47.54.+r, 89.75.--k}

\maketitle

Systems that are far from equilibrium exhibit remarkable physics: some
commonly known examples in the context of glassy systems \cite{review}
are ageing and metastability. There are, however,
much simpler consequences of nonequilibrium dynamics; in this Letter,
we present a model of cluster growth which is inherently disequilibrating.
Here, instead of available masses being equally distributed between
competing clusters, the largest cluster {\it always} wins. Somewhat surprisingly,the  two-step relaxation and ageing characteristic of glassy
dynamics is manifested for infinite systems. Additionally, in finite dimensions, the system relaxes asymptotically to a  metastable state
in a complex energy landscape; in each such state, a fraction of immortal
clusters from nontrivial spatial patterns.

In our model,
 $n$ pointlike, immobile clusters
with time-dependent masses
$m_i(t)$ for $i=1,\dots,n$,  
 evolve according to the following
 equations:
\beq
\dt{m_i}=\left(\frac{\a}{t}-\frac{1}{t^{1/2}}\sum_jg_{ij}\dt{m_j}\right)m_i
-\frac{1}{m_i}.
\label{dtm}
\eeq

These dynamical equations were originally written to model the
kinetics of black hole growth in a radiation fluid~\cite{archan,I}.
The present model is, however,
 of potential interest in many other situations; 
its dynamical behaviour in fact illustrates the
{\it rich-get-richer} principle
in economics~\cite{simon}, and its form bears some resemblance
to Lotka-Volterra type
 predator-prey models~\cite{murray} in biophysics.
In this Letter, we visualise Eqs. \ref{dtm} as
representing a simple model of aggregating
clusters in a  stirred colloidal system \cite{coll}.
 In our toy model, clusters grow by accreting
mass from the fluid and from other clusters; they can also dissolve
away if they are small, as the fluid is stirred. 
The gain term in (\ref{dtm}) is the sum of the free rate for an isolated cluster,
proportional to  the parameter~$\a>1/2$, and of constributions from all
other clusters via the fluid, with a 
 coupling $g_{ij}$ between clusters $i$ and $j$; the loss term is taken
to be inversely proportional to the cluster mass. The stirring
of the fluid results, in this simple-minded picture, in the explicit
time dependences in (\ref{dtm}).

We first recall \cite{archan, condmat} the one-cluster result.
(For convenience, 
we work with reduced time
$s=\ln\frac{t}{t_0}$ (to renormalise away the effect of
 initial time $t_0$),
 reduced masses 
$x_i=\frac{m_i}{t^{1/2}}$ and square masses $y_i=x_i^2=\frac{m_i^2}{t}$.)
Large clusters, whose initial mass $y_0$ is greater than some
threshold $y_\star$, 
with $y_\star (t_0)=\left(\frac{2t_0}{2\a-1}\right)$,
are {\it survivors:} they survive and keep on growing forever.
Small clusters, whose initial mass is below this threshold
dissolve and die out in a finite time. In a very dilute sediment,
we would thus expect globules of matter to be suspended indefinitely,
and to grow forever.

As the concentration of the solute is increased, interactions
between the colloidal clusters become significant. We consider
first two  interacting clusters with
 interaction strength $g$. If
 their initial masses
are exactly equal, this equality is maintained by symmetry forever.
The reduced mass  $x(s)$ of each mass then obeys:
\beq
x'=\frac{(2\a-1)x^2-2-gx^3}{2x(1+gx)}.
\label{dsx1}
\eeq
 The fixed points of \ref{dsx1}
 are given by
$(2\a-1)x^2-2-gx^3=0$; there is a critical value 
$g_c=\left(\frac{2(2\a-1)^3}{27}\right)^{1/2}$
which separates two kinds of behaviour. For large couplings $g>g_c$  , there is
 no fixed point; physically this implies that the clusters
feed on each other and disappear quickly. For small couplings
$g<g_c$, there are two positive fixed points
\beq
y_\star^{1/2}<x\one\hbox{ (unstable) }<(3y_\star)^{1/2}<x\de\hbox{ (stable)}.
\eeq
Small clusters, such that their mass $x_0<x\one$,
are dynamically attracted by  $x=0$; thus lighter clusters
dissolve away rapidly.
 Large clusters, such that their mass $x_0>x\one$, are dynamically
 attracted by $x\de$: heavier clusters thus grow forever
according to the law
as $m(t)\approx x\de t^{1/2}$. Here the interaction is small enough
that symbiosis is achieved; the clusters feed on each other so as to
increase, rather than deplete, mutual growth.

For two  clusters with initially unequal masses,
any small initial mass difference always diverges
exponentially at early times:
later stages of the dynamics cannot be described in closed form.
The details of the transient behaviour can be found in
 \cite{condmat}.
However, the asymptotic behaviour is such that the largest cluster wins:
{\it survival of the biggest is therefore the
 single generic scenario
for two clusters with unequal initial masses}.
The surviving large cluster may then either
also disappear in a finite time,
or survive and grow forever, depending on whether its
 mass at the time of the lighter cluster's death 
is less or greater than the threshold
 $y_\star$. This generalises easily to 
 any finite number $n\ge2$ of colloidal clusters; in every case, our model predits that
there will at most be one large sediment formed \cite{coll}.

To explore an infinite assembly of interacting colloidal clusters,
we explore the mean-field behaviour of the model; each cluster
is connected to every other by a weak
interaction
 $g=\frac{\gbar}{n}$. When
 $n\to\infty$ limit at fixed $\gbar$, the socalled thermodynamic limit,
one obtains \cite{condmat}:
\beq
y'(s)=\g(s)y(s)-2
\label{mfds}
\eeq
for the reduced square mass $y(s)$ of any of the clusters.
Eq.~(\ref{mfds}) can be solved only formally,
as it is self-consistent \cite{condmat}.

Things simplify considerably
when the rescaled coupling $\gbar$ is small; remarkably,
a {\it glassy} dynamics \cite{review, glassyrefs} with two-step relaxation is observed. 
In Stage~I, the clusters behave as if they were isolated: this
corresponds to {\it individual} behaviour. The survivors
of this stage are clusters whose initial masses exceed the 
threshold $y_\star$, exactly as in the one-cluster case recalled above.
In  Stage~II, all the survivors interact with each other;
the dynamics is thus {\it collective}, and turns out to be slow \cite{condmat}.
All but the largest cluster eventually die out during this stage, thus
in terms of experimental predictions \cite{coll}, again predicting a uniform
asymptotic sediment. This weakly interacting mean-field regime of our model
 of interacting colloids
clusters also exhibits other characteristic features
of glassy systems~\cite{glassyrefs} such as ageing; this
 originates in the presence
of two well-separated time scales of fast and slow dynamics,
whose ratio grows  as $1/\gbar^2$ \cite{condmat}.

 For an exponential distribution of initial masses,
in the late times of Stage~II, the survival probability  decays as
\beq
S(t)\approx\frac{2\a-1}{\gbar}\,\left(C\,\ln\frac{t}{t_0}\right)^{-1/2},
\label{stlate}
\eeq
with 
\beq
C=\pi
\label{cpi}
\eeq
irrespective of $\a$, ~$\gbar$ and the parameters of the exponential
distribution. Also,
 the mean mass of the surviving clusters grows as
\beq
\meansur{m}_t\approx\left(C\,t\,\ln\frac{t}{t_0}\right)^{1/2}.
\label{mtlate}
\eeq

The universality inherent in the scaling results~(\ref{stlate})--(\ref{mtlate})
is unusual, because it includes the prefactor $C$,
which is itself independent of the details of the initial distribution $P(y_0)$
of square masses.
It can in fact be shown  that $C$ only depends on the {\it tail exponent}
of this distribution in the vicinity of its upper bound~$y\max$, i.e.
whether the initial distribution of masses is bounded or not. The interested 
reader is
referred to \cite{condmat} for further details; here we simply point out that
this striking universality
is a major result of our present work. Also,
the logarithmic behaviour seen  is yet another signature
~\cite{glassyrefs}
of glassy dynamics. Returning to the physical system, this predicts
that in a very weakly interacting system of colloidal particles, there is
a slow freezing in of fluid disorder\cite{coll}, resulting in a uniform
sediment at asymptotically long times.

Finally, we study a lattice version of the model
( we choose the chain ($D=1$), the square lattice ($D=2$), and the cubic lattice ($D=3$))
in order to probe the effect of fluctuations.
Clusters now sit on the vertices $\n$ of a regular lattice,
with nearest-neighbour interaction
 $g$.
In the limit of
 weak coupling ($g\ll1$), our dynamical equations are:
\beq
x'_\n=\left(\frac{2\a-1}{2}+g\sum_{\m}\left(\frac{1}{x_\m}-\a x_\m\right)
\right)x_\n-\frac{1}{x_\n},
\label{dsg}
\eeq
where $\m$ runs over the $z=2D$ nearest neighbours of site $\n$,

The dynamics generated by~(\ref{dsg}),
again consists of two successive well-separated stages.
 Fast individual dynamics are exhibited in Stage~I, where
the mass of each cluster evolves as if it were isolated.
As in the mean-field case, the survival probability $S(s)$ decays rather fast
from $S(0)=1$ to a plateau value $S\one$.
The effects of going beyond mean field are only palpable
in Stage II, where interactions become relevant and lead to a slow
dynamics which
is now very different from the mean-field scenario.
The survival probability $S(s)$ in fact decays
from its plateau value $S\one$ to a non-trivial limiting value $S\infy$,
reflective of metastability. The effect of fluctuations is already palpable:
unlike the mean field result, this predicts that a {\it finite} number of 
clusters will survive. We elaborate on this below.

\begin{figure}[htb]
\begin{center}
\includegraphics[angle=90,width=.6\linewidth]{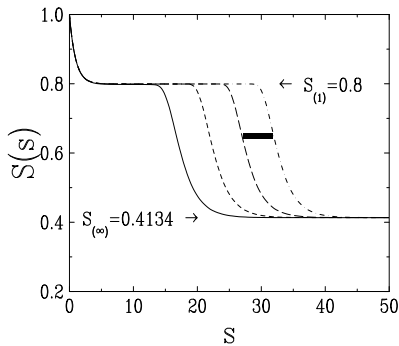}
\caption{\small
Plot of the survival probability $S(s)$ on the chain with $S\one=0.8$.
Left to right:
Full line: $g=10^{-3}$.
Dashed line: $g=10^{-4}$.
Long-dashed line: $g=10^{-5}$.
Dash-dotted line: $g=10^{-6}$.
The thick bar has length $2\,\ln 10=4.605$~(see text).}
\label{figc}
\end{center}
\end{figure}

Consider~(\ref{dsg}) for two neighbouring clusters $\n$ and $\m$
which have both survived Stage~I.
The contribution of cluster $\m$ to the large parenthesis in the
right-hand side of~(\ref{dsg}) is proportional to~$\a gx_\m$.
In the absence of coupling,
we have $x_\m\sim\e^{(2\a-1)s/2}$ \cite{archan, I}.
The characteristic time scale of Stage~II, that is the time
at which interactions become significant,
is reached when the product~$gx_\m$ becomes of order unity:
\beq
s_c\approx\frac{2}{2\a-1}\,\ln\frac{1}{g},
\label{sc}
\eeq
i.e. $t_c\sim t_0\,g^{-2/(2\a-1)}$.
Thus, in this weak-coupling limit,
the separation of time scales between fast individual
and  slow collective dynamics
is
 parametrically large.
Figure~\ref{figc}
illustrates this two-step relaxation in the decay of the
 survival probability $S(s)$ in one dimension.
Both stages of the dynamics appear clearly on the plot;
different values of the interaction cause the system
to {\it age} differently. In each case,
a plateau  is reached at $S\one=0.8$; however, the weaker
the interaction, the longer the system takes to reach the
asymptotic state, which occurs at
a non-trivial limit survival probability
$S\infy\approx0.4134$. Since each curve
corresponds to a  
decade shift in
 interaction strength $g$, it
is shifted in terms of the onset time $s_c$ of slow dynamics by~$2\,\ln 10$ (thick bar),
in excellent agreement with the estimate~(\ref{sc}).

At the end of Stage~II,
the system is left in a non-trivial {\it attractor},
which consists of a pattern where each cluster is {\it isolated}:
it is therefore a {\it survivor} and keeps growing forever.
We call these attractors {\it metastable states},
since they form valleys in the existing random energy landscape;
 the particular metastable state chosen
by the system 
(corresponding to a particular choice of pattern) is the one which
can most easily be reached in this landscape\cite{review, tap, glassyrefs}.
The number~$\N$ of these states
generically grows exponentially with the system size (number of sites) $N$:
\beq
\N\sim\exp(N\Sigma).
\eeq
with $\Sigma$  the configurational entropy
or {\it complexity}.

The limit survival probability $S\infy$ (Figure~\ref{figc})
is just the density of a typical attractor,
i.e., the fraction of the initial clusters which survive forever.
It
 obeys the inequalities
\beq
S\infy\le S\one,\qquad S\infy\le1/2.
\eeq
The first inequality expresses the fact  that clusters generically disappear:
the difference $1-S\one$ (resp.~$S\one-S\infy$) is the fraction of clusters
which are dissolved out during Stage~I (resp.~Stage~II).
The second inequality is a consequence of the fact that
each surviving cluster is isolated: the densest configuration
for which this is the case is when
 either of the two sublattices is occupied, at which point the density
is exactly $1/2$.
This value $1/2$ of the highest density holds for the large family
of so-called {\it bipartite} lattices,
which includes the hypercubic lattices we have considered here (but does not,
for example, include the triangular lattice).

For a given class of initial mass distributions,
the limit survival probability $S\infy$
is a monotonically increasing function of the plateau value $S\one$;
the more the number of survivors after
Stage I, the higher will evidently be the number of {\it immortal} clusters.
For $S\infy=0$, $S\one$ is trivially $0$;
as the  $S\one\to1$ limit is approached, the
 non-trivial maximum value $S\infymax<1/2$
is reached. Additionally, when $S\one$ is small,
it can be shown that $S\infy$ is also small,
and that it depends on $S\one$ alone.
This is shown below.

We define a 
 {\it supercluster} as a set of $k\ge1$ connected clusters
which have survived Stage~I,
 such that all their neighbours have disappeared during Stage~I.
The fate of superclusters depends on their size $k$ as follows.
\begin{itemize}
\item[$\star$] $k=1$:
If a supercluster consists of a single isolated cluster,
it is a survivor, because its mass exceeds the survival threshold $y_\star$.
For $z=2D$ and independently of initial mass distributions,
a supercluster with $k=1$ occurs 
with density $p_1=S\one(1-S\one)^{2D}$. This
corresponds to the survival of one cluster after, and the disappearance
of its $2D$ neighbours during, Stage I.
\item[$\star$] $k=2$:
If a supercluster consists of a pair of neighbouring clusters
(represented as~\bl\bl)
both clusters evolve according to the dynamics described above:
the smaller dies out, while the larger is a survivor.
We are thus left with~{\bl\vi} or~{\vi\bl} in the late stages of the dynamics.
Such an event takes place with density $p_2=S\one^2(1-S\one)^{2(2D-1)}$.
\item[$\star$] $k\ge3$:
If three or more surviving clusters form a supercluster,
they may a priori have more than one possible fate.
Consider for instance a linear supercluster of three clusters (\bl\bl\bl).
If the middle one disappears first (\bl\vi\bl),
the two end ones are isolated, and both will be survivors.
If one of the end ones disappears first (e.g.~\bl\bl\vi),
the other two form an interacting pair,
and only the larger of those two will survive forever (e.g.~\bl\vi\vi).
The pattern of the survivors,
and even their number, therefore {\it cannot} be predicted a priori.
\end{itemize}
The above enumeration implies $S\infy=p_1+p_2/2+\cdots$,
where the dots stand for the unknown contribution of superclusters
with $k\ge3$.
As $p_1\sim S\one$, $p_2\sim S\one^2$, and so on, we are left with the expansion
\beq
S\infy=S\one-D\,S\one^2+\cdots
\label{expan}
\eeq
The dependence of $S\infy$ on details of the initial mass distribution
at fixed $S\one$ therefore only appears at order $S\one^3$. These results
apply to the dilute limit, when few clusters survive stage I.

In the limit that there are many clusters which survive
the fast dynamics of stage I, i.e.
 $S\one\to1$, the limit survival probability, as mentioned above,
reaches a non-trivial maximum value $S\infymax<1/2$. This
 depends very weakly on the mass distribution;
for instance, in one dimension one has
$S\infymax\approx0.441$ for an exponential distribution
and $S\infymax\approx0.446$ for a uniform distribution.
We present below a way of visualising these dense distributions
of immortal clusters.

If ($S\infy=1/2$) on, say, a square lattice,
(i.e. the highest density of immortal
clusters is reached),
there are only two possible `ground-state' configurations of the system.
These correspond to the full occupation of one of the two sublattices,
with its counterpart being completely empty.
In this limit, the two possible
patterns of immortal clusters are each perfect checkerboards of
one of two possible parities.

At every site $\n$
with integer co-ordinates $(n_1,n_2,\dots,n_D)$ we define the {\it survival index}
\beqa
\s_\n&=&1\quad\hbox{if the cluster at site~$\n$ is a survivor}\nonumber\\
     &=&0\quad\hbox{else} 
\eeqa

and the {\it checkerboard index}
\beq
\phi_\n=(-1)^{\s_\n+n_1+\cdots+n_D}.
\eeq
The survival index depicts very simply the pattern of surviving clusters
surrounded by empty sites:
The checkerboard index, on the other hand, represents,
for each site, the local choice of one of the two symmetry-related
`ground states', i.e., of one of the two sublattices.
This is easiest to understand using a one-dimensional example:
the two ground states are $+-+-+\cdots$ or $-+-+-\cdots$
All the $\phi_n$ are equal to $-1$ in the first pattern,
and equal to $+1$ in the second pattern.
The checkerboard index $\phi_\n$ thus classifies
each site according to the {\it parity} of the particular ground state selected locally
at this site.

\begin{figure}[htb]
\begin{center}
\includegraphics[angle=0,width=.35\linewidth]{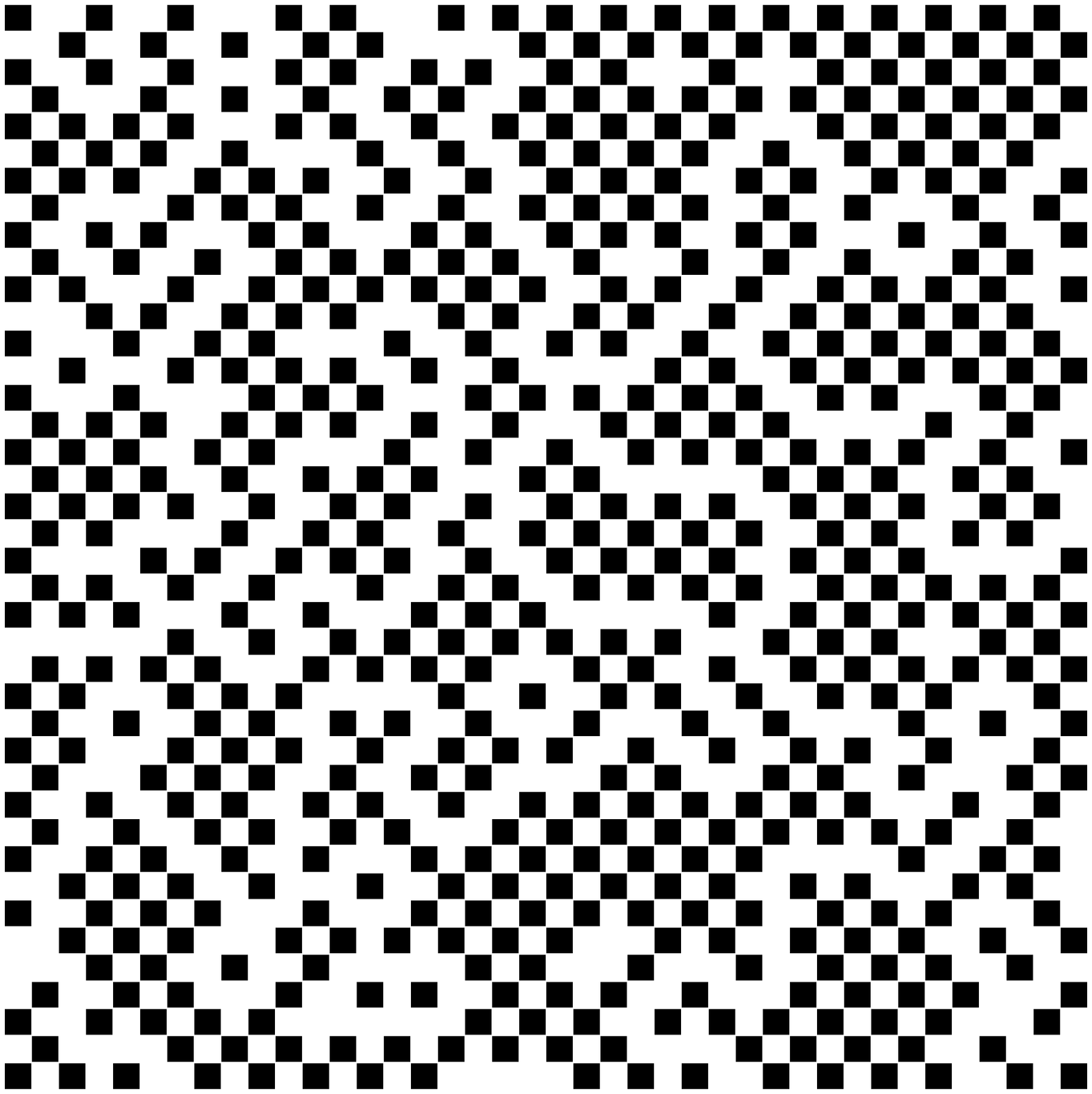}
{\hskip 1mm}
\includegraphics[angle=0,width=.35\linewidth]{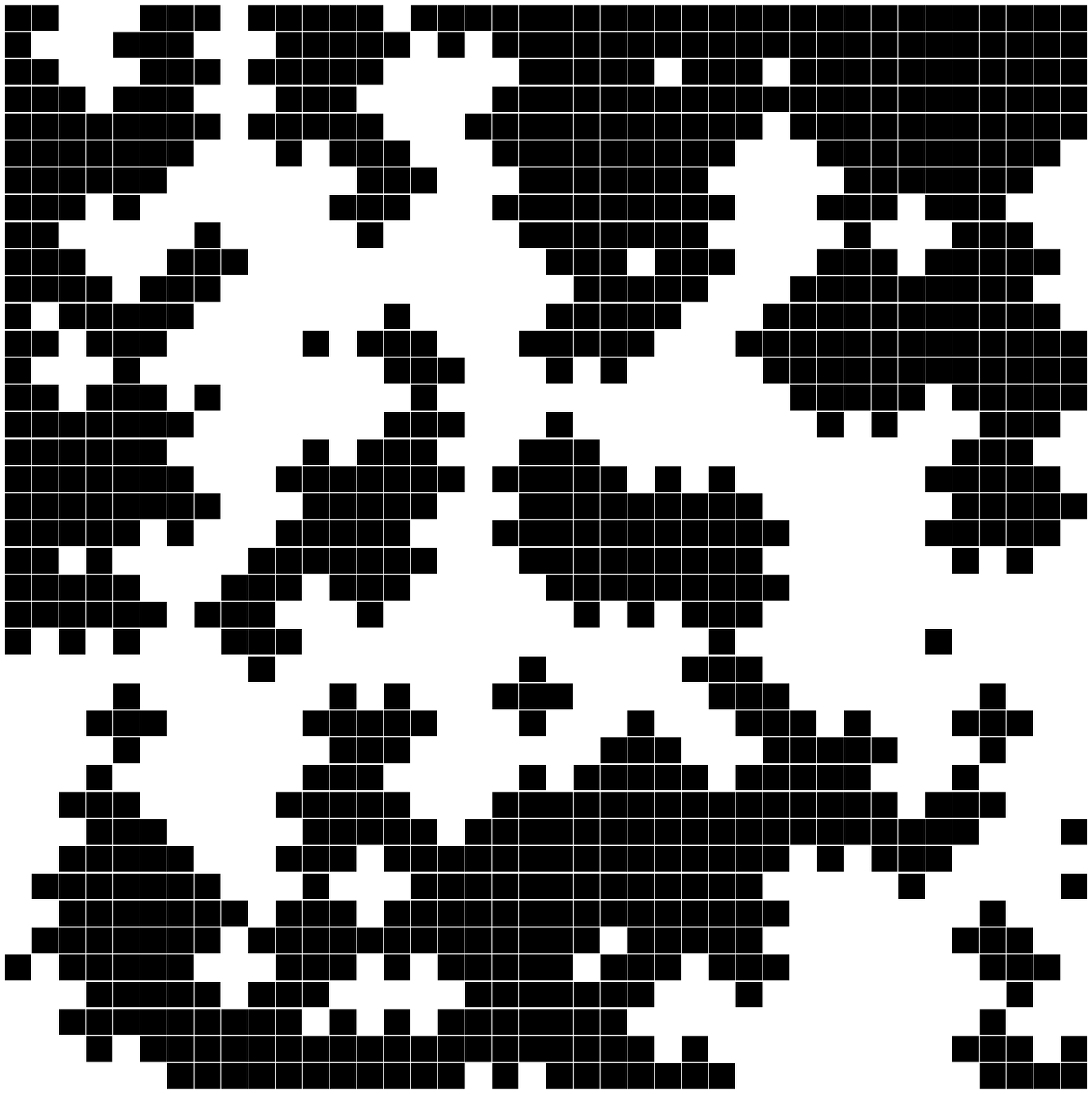}
\caption{\small Two complementary representations
of a typical pattern of surviving clusters on a $40^2$ sample of
the square lattice,
with $S\one=0.9$, so that $S\infy\approx0.371$.
{\bf Left}: Map of the survival index.
Black  squares represent immortal sites for which $\s_\n=1$,
while white squares represent dead sites for which
$\s_\n=0$.
{\bf Right}: Map of the checkerboard index.
Black squares represent $\phi_\n=+1$, while white squares represent
$\phi_\n=-1$.}
\label{fige}
\end{center}
\end{figure}

For generic initial conditions, i.e. a random distribution of initial masses,
the immortal sites will evidently not form a perfect checkerboard. However,
if the initial masses are large enough, the number of survivors of Stage I
dynamics will be large, and the corresponding
survival probability $S\one$ after Stage~I  close to unity. In this
limit,
the asymptotic survival probability $S\infy$ will be close to
 its `ideal' value of~$1/2$.
The resulting pattern will now exhibit a {\it local} checkerboard structure,
with frozen-in defects between patterns of different parities; the
 random structure of defects is of course entirely inherited
from the (random) initial mass distribution,
since the dynamics is fully deterministic. This is evident
from 
Figure~\ref{fige}, which shows a map of the survival index
and of the checkerboard index
for the same attractor for a particular sample of the square lattice.
The immortal (black) clusters in the left-hand part of the figure
are surrounded by rivulets of voids, which are a consequence
of initial conditions; in the right-hand figure, the deviation
from a perfect checkerboard structure (all black or all white)
is made clearer. The patterns make it clear that neighbouring
sites must be fully anticorrelatedm
because each immortal cluster is surrounded by voids.
However, at least close to the limit $S\infy=1/2$,
immortal sites are very likely to have  next-nearest neighbours 
which are likewise immortal and massive.
The detailed examination of survival and mass correlation functions
made in a longer paper \cite{condmat} confirms these expectations.

To conclude,
 we have presented in this Letter a very simple 'winner-takes-all' model
 of cluster aggregation
in stirred colloidal systems. Both mean-field and finite dimensional
explorations of this model show a striking and a priori unexpected glassy
phase; the system of interacting clusters shows {\it ageing} until it reaches
its asymptotic state. The inclusion of fluctuations in the model via
a finite-dimensional approach causes the replacement of the somewhat
staid mean-field behaviour (which predicts a uniform sediment composed
of one cluster at most) by something far more exciting; a random-energy
landscape emerges, with many possible minima as its metastable states, and the
system descends to the most accessible one. Each such metastable state
is a complex pattern of isolated clusters, each of which, by virtue of its isolation, is {\it immortal}.

\end{document}